\documentclass{ws-procs10x7}
\usepackage{balance}
\usepackage{psfig}

\newcolumntype{d}[1]{D{.}{.}{#1}}

\newcommand{\ee}{$e^+e^-~$}
\newcommand{\GG}{$\gamma\gamma~$}
\newcommand{\pp}{$p\bar{p}~$}
\newcommand{\dsf}{$F_{jj}^D$}
\newcommand{\rsn}{$R_{SD}^{ND}$}
\newcommand{\xbj}{$x_{Bj}$}
\newcommand{\qs}{$Q^2$}
\newcommand{\et}{$E_{T}$}
\newcommand{\etal}{\emph{et al.}~}
\newcommand{\fhf}{$F_{HF}$}

\makeindex
\begin{document}

\title{DIFFRACTION AT THE TEVATRON}

\author{A. HAMILTON$^*$ }

\address{DPNC, University of Geneva, Geneva, CH-1211, Switzerland, ahamil@fnal.gov\\$^*$On behalf of the CDF and  D\O\ Collaborations}




\twocolumn[\maketitle\abstract{This article contains a summary of the recent results in diffractive physics at the Tevatron.  Results from the CDF diffraction program include the single diffractive to non-diffractive ratio in dijet events, observation of exclusive \ee production via two-photon exchange, hints of exclusive \GG production via Double Pomeron Exchange (DPE), and the observation of exclusive dijet production via DPE.  A summary of the current plans for the D\O\ diffractive physics program is also presented.}
]

\section{Introduction}
Diffractive \pp collisions are mediated by the exchange of a colorless object with the quantum numbers of the vacuum, the Pomeron.   Pomeron exchange allows the $p$ and/or $\bar{p}$ to stay intact and produces a rapidity gap (a region of rapidity with no particles).  Both the CDF and D\O\ Collaborations have extensive diffractive physics programs.  Recent results from the CDF Collaboration are presented along with the diffractive physics plans of the  D\O\  Collaboration.

\section{Results from CDF}\label{sec:cdf}
The CDF diffractive physics program can be broken into two broad categories: inclusive diffraction and exclusive production.  The primary goal of the inclusive diffraction program is to understand the nature of the Pomeron exchange and study the ``diffractive structure function''\cite{dsf} (\dsf).  The primary motivation for studying exclusive production at the Tevatron is to test the feasibility of using exclusive production to search for and study the Higgs boson and other new physics at the LHC\cite{kmr}.  In leading order QCD, exclusive production occurs through gluon-gluon fusion, while an additional soft gluon screens the color charge allowing the protons to remain intact\cite{kmr}.  Exclusive production of dijets, \GG, and Higgs bosons are possible through this mechanism.   Exclusive production can also occur through two-photon exchange: $\gamma\gamma \rightarrow \ell^+\ell^-$.

\subsection{Inclusive Diffraction Results}
\begin{figure}[b]
\centerline{\includegraphics[width=2.2in]{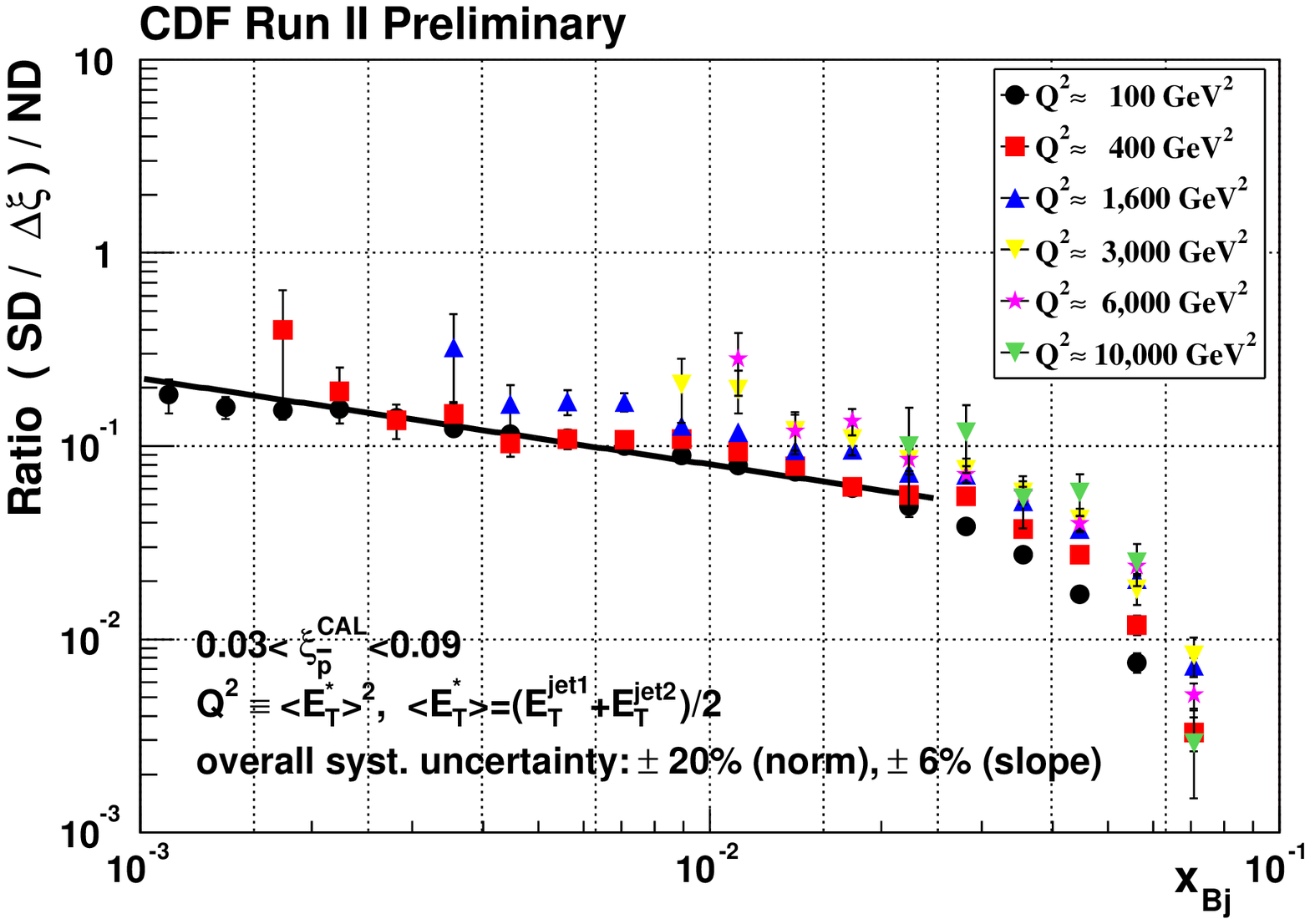}}
\centerline{\includegraphics[width=2.2in]{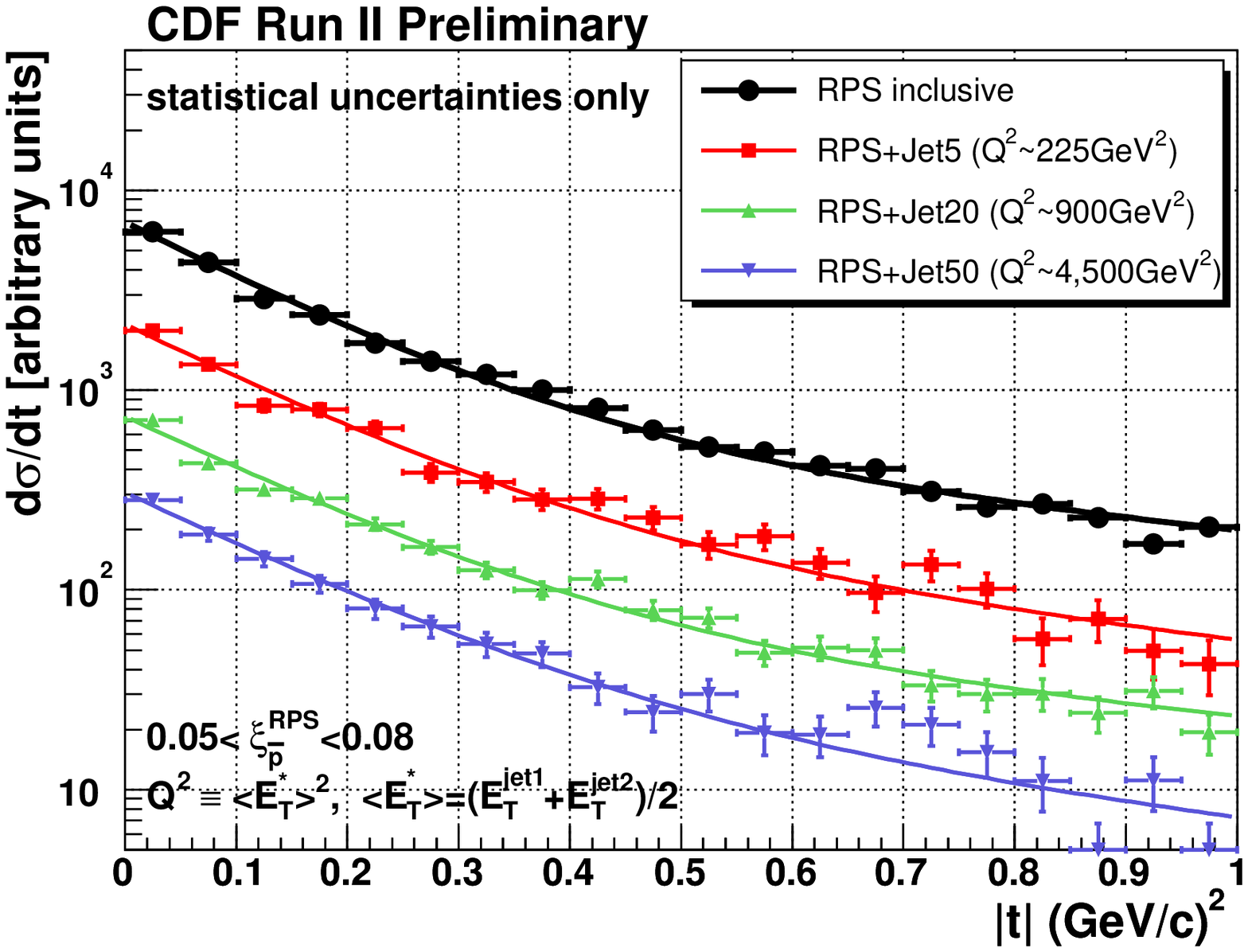}}
\caption{(top) \rsn as a function of \xbj for different \qs.  (bottom) SD dijet distribution as a function of $t$ for different \qs.}
\label{fig:rsn}
\end{figure}

Information about \dsf~ relative to the proton PDF can be extracted by examining the ratio of single-diffractive (SD) to non-diffractive (ND) dijet event rates (\rsn).  The SD data sample is collected by triggering on the leading antiproton in the roman pot spectrometer\cite{fwddet} plus at least one jet in the central detector.  The ND data sample is collected using just the central jet trigger.  The ratio of the SD to ND dijet production is shown in Fig.~\ref{fig:rsn} (top) as a function of the Bjorken-$x$ (\xbj).  The figure shows \rsn ~for different values of \qs, where \qs~is defined as the square of the mean dijet \et.  In the range \mbox{$100<Q^2<10000~GeV^2$} there is no significant \qs ~dependence observed, indicating that the evolution of the Pomeron could be similar to that of the proton.

The \qs ~dependence of SD dijet events is also examined as a function of $|t|$, the squared momentum transfer.  Figure~\ref{fig:rsn} (bottom) shows that the slope (in arbitrary normalization) of the $t$-distribution at low $|t|$ is independent of \qs.  Measurement of the value of the slope is currently under way.

\subsection{Exclusive \ee Production}
\begin{figure}[b]
\centerline{\includegraphics[width=2.2in]{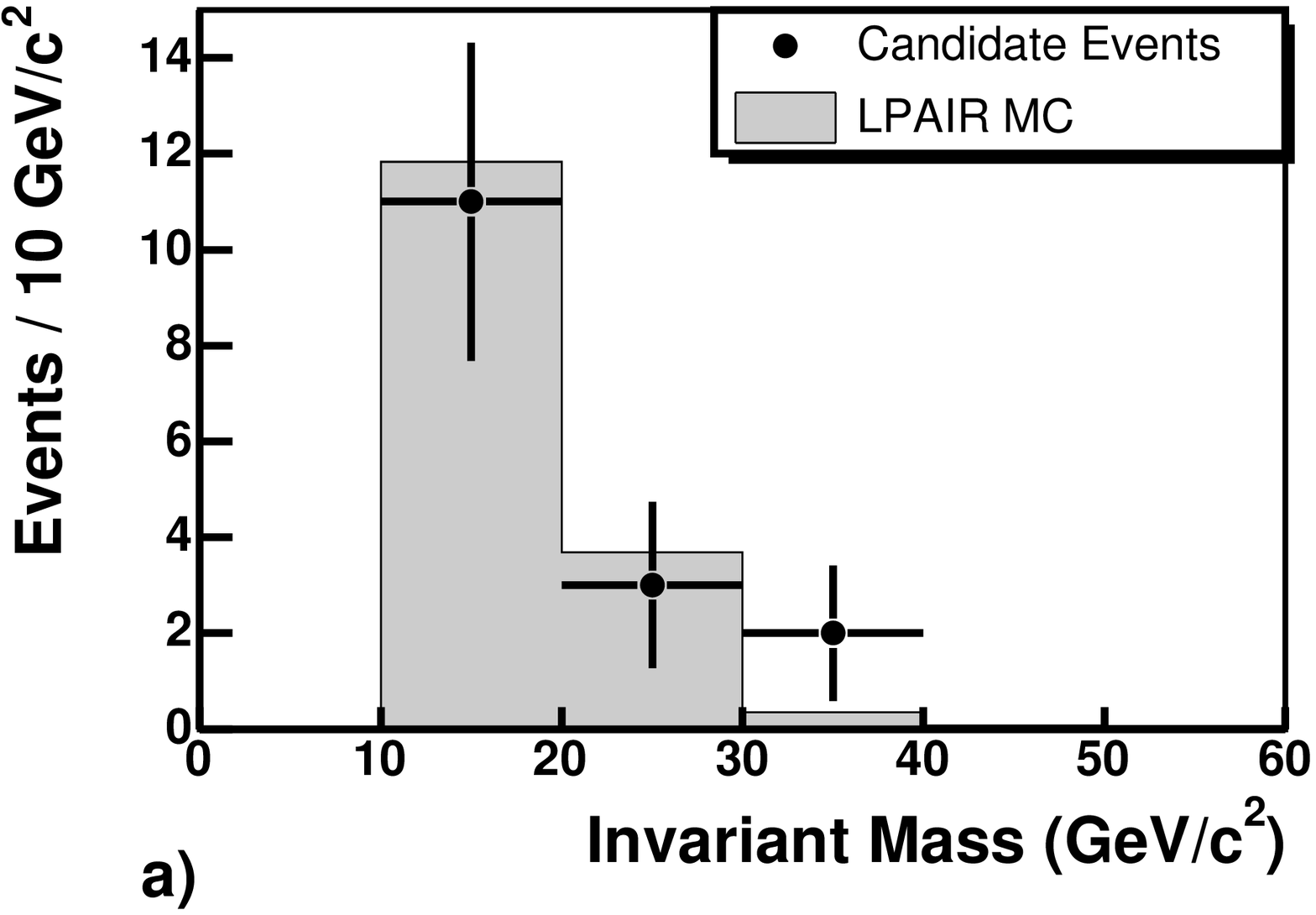}}
\centerline{\includegraphics[width=2.2in]{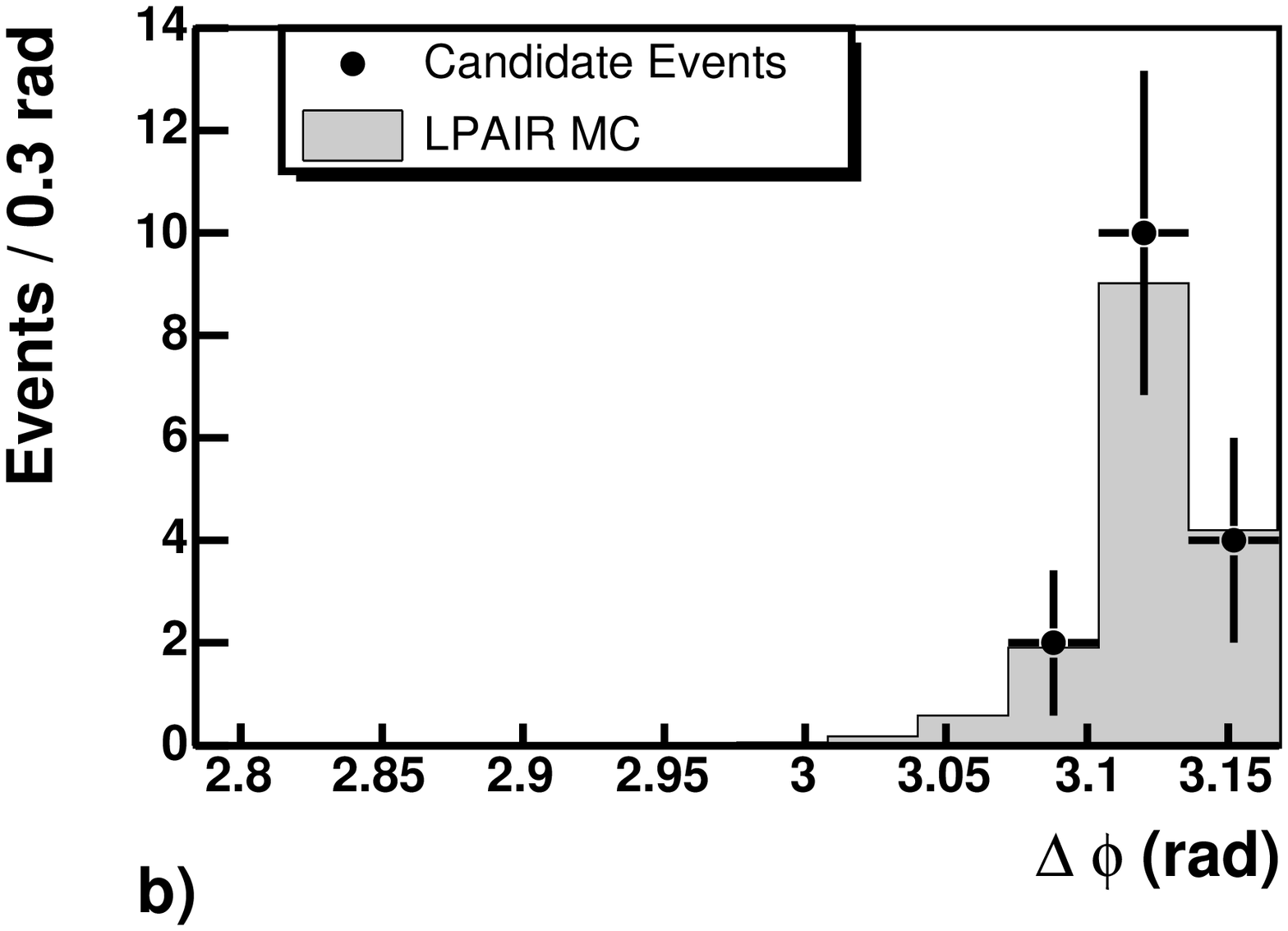}}
\caption{a) Invariant mass of the \ee system. b) $\Delta \phi$ between the $e^+$ and $e^-$.}
\label{fig:ee}
\end{figure}

CDF has observed exclusive \ee pro\-duc\-tion via two-photon exchange ($p+\bar{p} \rightarrow p+\gamma\gamma+\bar{p} \rightarrow p+e^+e^-+\bar{p}$) for the first time in hadron-hadron collisions.  Events were selected with a trigger requiring two electromagnetic (EM) calorimeter clusters and a veto on the activity in the beam shower counter detectors\cite{fwddet}.   Further offline selection criteria are applied requiring the two EM clusters to be consistent with electrons (\mbox{\et$>5~GeV$} and \mbox{$|\eta|<2.0$}) and no activity other than the electrons observed in \mbox{$|\eta|<7.4$}.  The outgoing protons are not observed.  There are 16 candidate events observed in an integrated luminosity of \mbox{$532 \pm 32 ~\mathrm{pb}^{-1}$}.   The background estimate of \mbox{$1.9\pm0.3$} events includes backgrounds from dissociation (two-photon exchange events in which one or both of the outgoing protons dissociate), inclusive processes (like Drell-Yan), hadronic jets faking the electrons, and cosmic rays.  The kinematic distributions are shown to agree with the predictions from {\sc lpair}\cite{lpair} Monte Carlo (MC) in Fig.~\ref{fig:ee}.  The measured cross section of \mbox{$1.6 ^{+0.5}_{-0.3}\mathrm{(stat)}\pm0.3\mathrm{(syst)}~\mathrm{pb}$} agrees with the theoretical cross section \mbox{$1.711\pm0.008$~pb} given by {\sc lpair}.  

\subsection{Exclusive \GG Production}
Three exclusive \GG candidate events are found by following the same selection criteria as the exclusive \ee production analysis, except that rather than requiring the EM cluster to have a track, the EM cluster is required to have either no matched tracks or a pair of tracks consistent with a $\gamma \rightarrow $~\ee conversion.  A cross section for this process is not yet available because the background estimates are not complete. The exclusive production estimates of Khoze~\etal\cite{kmr}, as implemented in the {\sc exhume}\cite{exhume} MC, predict \mbox{$1^{+3}_{-1}$}~exclusive \GG events in the data sample.  

\subsection{Exclusive Dijet Production}
CDF has observed exclusive dijet production in \pp collisions.  A sample of Double Pomeron Exchange (DPE) dijet data is collected by requiring the leading $\bar{p}$ to be detected in the roman pot spectrometer, a veto on the detector activity in \mbox{$3.6<\eta<5.9$} (on the $p$ side), plus at least two jets with \et~$>5$~GeV.  The data sample is examined using the mass of the dijet system $M_{jj}$ divided by the mass of the entire central system $M_x$: \mbox{$R_{jj}=M_{jj} / M_x$}.  

\begin{figure}[b]
\centerline{\includegraphics[width=2.2in]{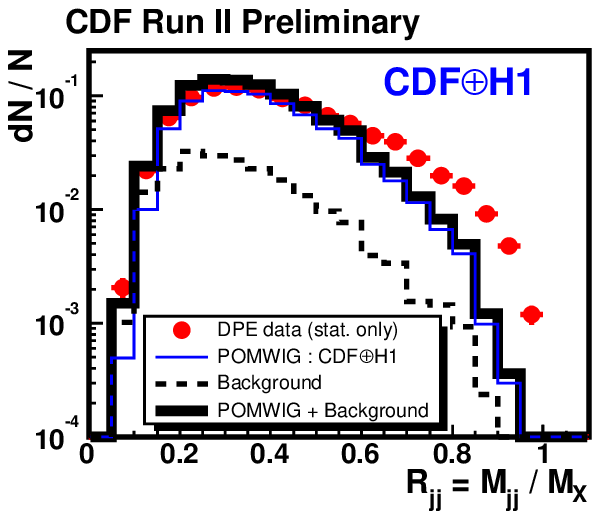}}
\centerline{\includegraphics[width=2.2in]{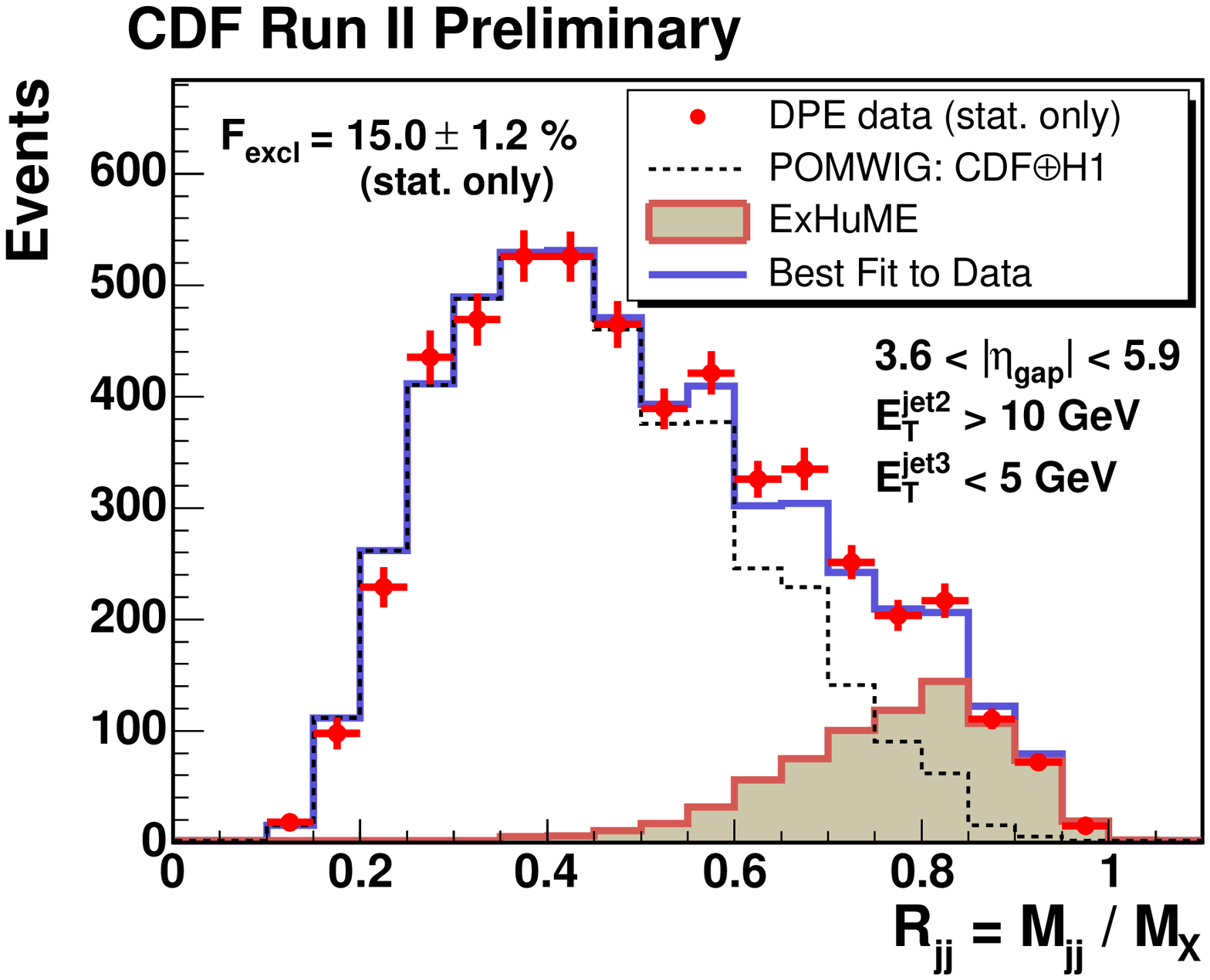}}
\caption{(top) $R_{jj}$ distribution for the DPE dijet sample and its comparison to {\sc pomwig}.  (bottom) $R_{jj}$ distribution of the subsample selected to enhance the exclusive component and its fit to a combination of {\sc pomwig} and {\sc exhume}. }
\label{fig:excdijet}
\end{figure}

Three MC programs are used to model the data.  {\sc exhume}\cite{exhume} and {\sc dpemc}\cite{dpemc} model exclusive interactions, while {\sc pomwig}\cite{pomwig} models non-exclusive diffractive interactions.  A comparison of this data sample to {\sc pomwig} is shown in Fig.~\ref{fig:excdijet}(top) as a function of $R_{jj}$.  The exclusive component of the dataset is enhanced by requiring both jets to have \et~$>10$~GeV and a veto on any additional jets with \et~$>5$~GeV.  This enhanced dataset is fit to a combination of {\sc pomwig} and {\sc exhume}, shown in Fig.~\ref{fig:excdijet}(bottom).  A combination of {\sc dpemc} and {\sc pomwig} also fit well with this subsample.  

\begin{figure}[b]
\centerline{\includegraphics[width=2.2in]{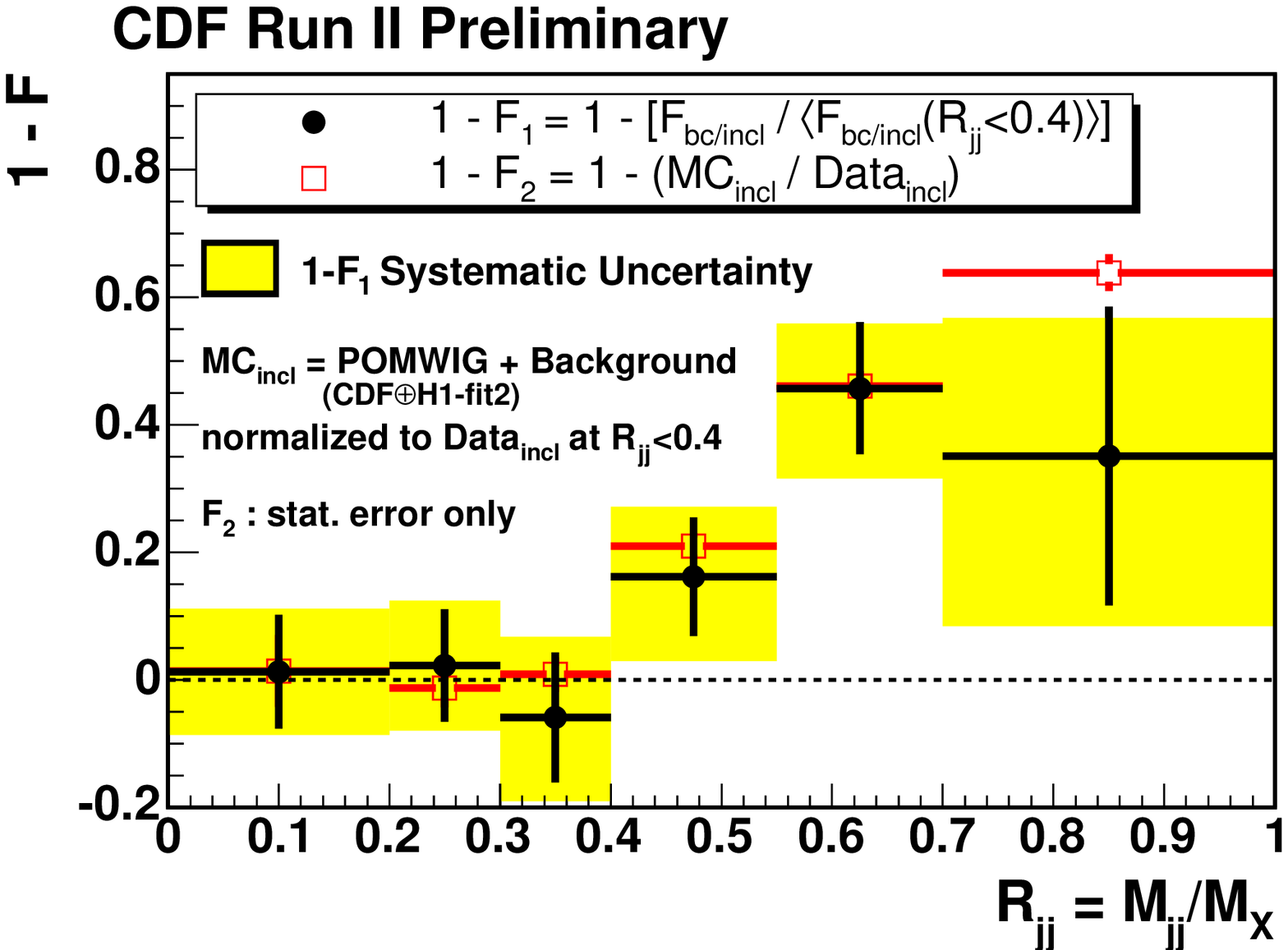}}
\centerline{\includegraphics[width=2.2in]{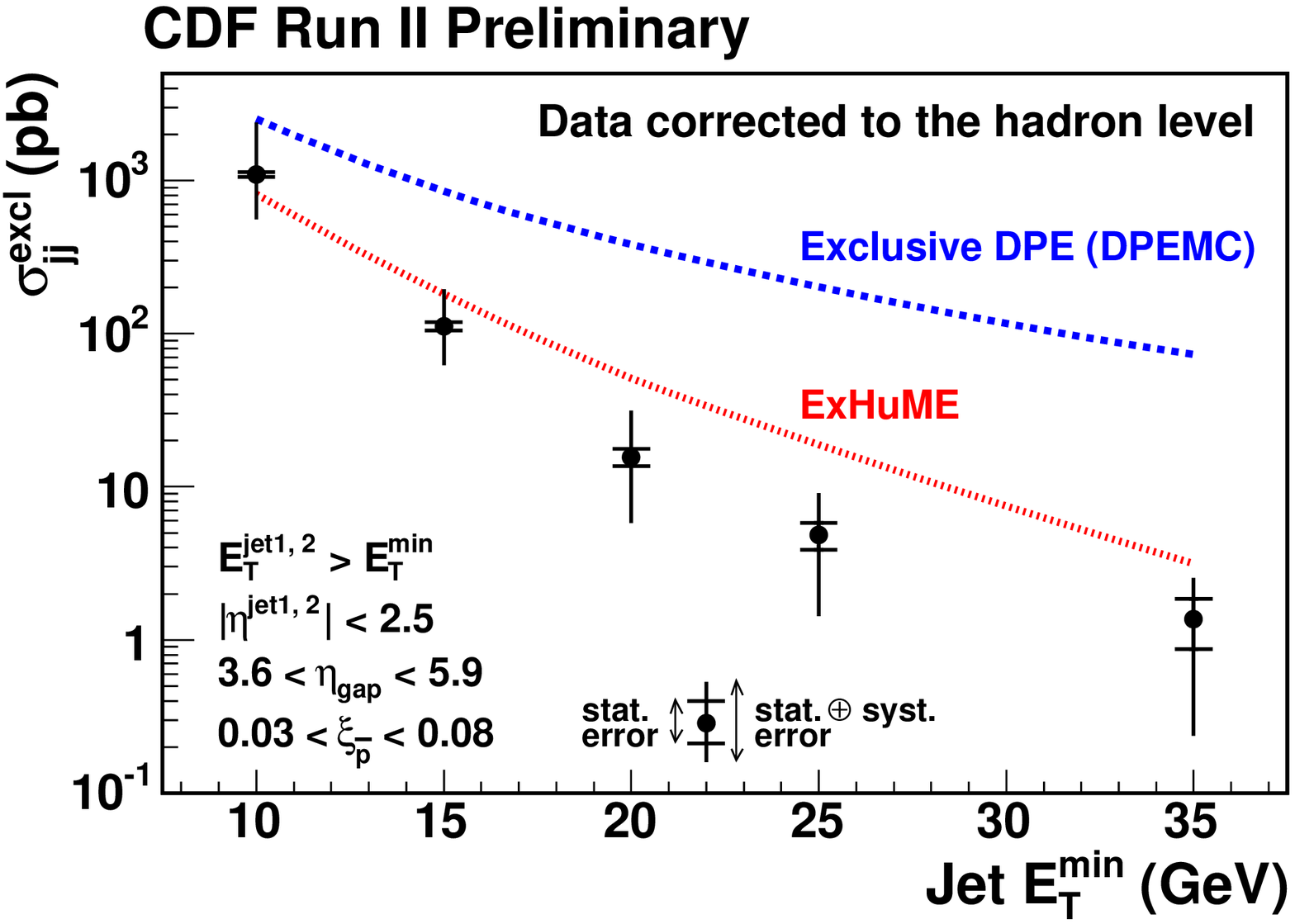}}
\caption{(top) \fhf vs. $R_{jj}$ as obtained directly from the DPE heavy flavor study and  as expected from the inclusive DPE dijet study.  (bottom) Measured exclusive dijet production cross section compared to {\sc dpemc} and {\sc exhume} predictions.}
\label{fig:sup}
\end{figure}

Another observable available to measure the contribution from exclusive events in the DPE dijet sample is the fraction of heavy flavor quark jets \fhf ~in the sample.  The \fhf ~is expected to be suppressed at high $R_{jj}$ due to the $J_Z=0$ spin selection enhancement in exclusive production\cite{kmr}.  Figure~\ref{fig:sup}(top) shows that this suppression is found in the data and compares it to the expectations from the previous study.  The figure shows that the two approaches agree.  The cross section of exclusive dijet production has been measured by multiplying the cross section of the DPE sample by the fraction of exclusive events in the sample and accounting for acceptances.  The cross section is compared to {\sc dpemc} and {\sc exhume} in Fig~\ref{fig:sup} which shows that the result favors the {\sc exhume} predictions.

\section{Diffractive Physics Plans at D\O\ }
The D\O\ detector is equipped with forward proton detectors\cite{fpd} and the D\O\ Collaboration is actively investigating many aspects of diffractive physics.  The studies in progress include:  inclusive DPE, measurement of \dsf, DPE with jets, diffractive production of heavy flavor, diffractive production of W/Z, and exclusive production.  While no new results are available at this time, results are expected in the near future.


\balance

\end{document}